\documentclass [times,10pt,twocolumn]{article}
\usepackage[utf8]{inputenc}
\usepackage{main}
\usepackage{algorithm,algpseudocode}
\usepackage{multirow}
\usepackage{subcaption}
\usepackage{ulem}
\usepackage{colortbl}
\usepackage{graphicx}
\usepackage{authblk} 
\usepackage{enumitem}
\newtheorem{Definition}{Definition}
\author[1]{Kendra M.L. Cooper}
\author[2]{Hassan Khosravi}
\affil[1]{Independent Scholar, Canada}
\affil[2]{University of Queensland, Australia}

\begin{document}
\title{Multilevel Visualisation of Topic Dependency Models for Assessment Design and Delivery: A Hypergraph Based Approach}
\date{April 2019}

\maketitle

\begin{abstract}
The effective design and delivery of assessments in a wide variety of evolving educational environments remains a challenging problem. Proposals have included the use of learning dashboards, peer learning environments, and grading support systems; these embrace visualisations to summarise and communicate results. In an on-going project, the investigation of graph based visualisation models for assessment design and delivery has yielded promising results.
Here, an alternative graph foundation, a two-weighted hypergraph, is considered to represent the assessment material (e.g., questions) and their explicit mapping to one or more learning objective topics. The visualisation approach considers the hypergraph as a collection of levels; the content of these levels can be customized and presented according to user preferences. A case study on generating hypergraph models using commonly available assessment data and a flexible visualisation approach using historical data from an introductory programming course is presented. \let\thefootnote\relax\footnotetext{DOI reference number: 10.18293/DMSVIVA2019-018}
\end{abstract}

\section{Introduction}
Assessment remains a core educational activity, even as environments continue to evolve beyond the traditional classroom. Blended, flipped, and massive open on-line courses are supported by a wide range of assessment tools and techniques \cite{Englund2017}. Instructors have many options for assessing the required topics of a course
in a particular offering
\cite{Griffin2015}. More traditional material includes homework assignments and examinations; emerging material includes question repositories and games. This rich variety also introduces new challenges to educational stakeholders (e.g., (e.g., students, instructors, administrators, education researchers) with respect to evaluating the coverage of assessment material and communicating achievements. Students may find it challenging to infer their strengths and weaknesses with respect to the topics and their relationships, which can impede their studies. Outside a classroom program administrators, course designers, course co-ordinators, and researchers also face challenges. Administrators find it challenging to compare the content and difficulty of formal assessments as well as students' outcomes across different offerings of a course. Course designers and co-ordinators find it challenging to ensure the required topics and their relationships (e.g. questions with a combination of topics $a$, $b$ and $c$) have been assessed. Educational researchers need to compare the achievement results between control and experimental groups. 
 
To address these assessment challenges, a new field, known as ``Learning Dashboards" has emerged, which embraces learning analytics and educational data mining \cite{schwendimann2016perceiving,Kirschner2012,Park2015}. These dashboards help users to interactively explore and understand data sets through analysis and visualisation techniques. A variety of traditional plots and charts have been adopted (e.g., pie, box, histogram, radial) to visualise the achievements of students. The research typically considers independent (stand-alone) topics. Additional discussion on the related work is presented in Section \ref{related-work}.

As part of an on-going project, the authors of this paper have explored a collection of topic dependency models (TDMs) for assessment in which the relationships among topics are considered \cite{Khosravi2018, Cooper2018}. The TDMs use a two-weighted undirected graphs foundation to formally represent and visualise a wide variety of assessment data for one or more topics (i.e., topics and their dependency relationship) to meet the needs of diverse stakeholders. The collection consists of a course reference model (to establish the topics and dependencies covered in a course), in addition to classroom models, both static and dynamic. As the work presented in this article leverages these results, a background section on the orignal TDM collection is presented in Section \ref{sec:background}. 

\label{sec:background}
\begin{figure*}[!htb]
		\centering
        \includegraphics[width=1.0\textwidth,height=11.5cm]{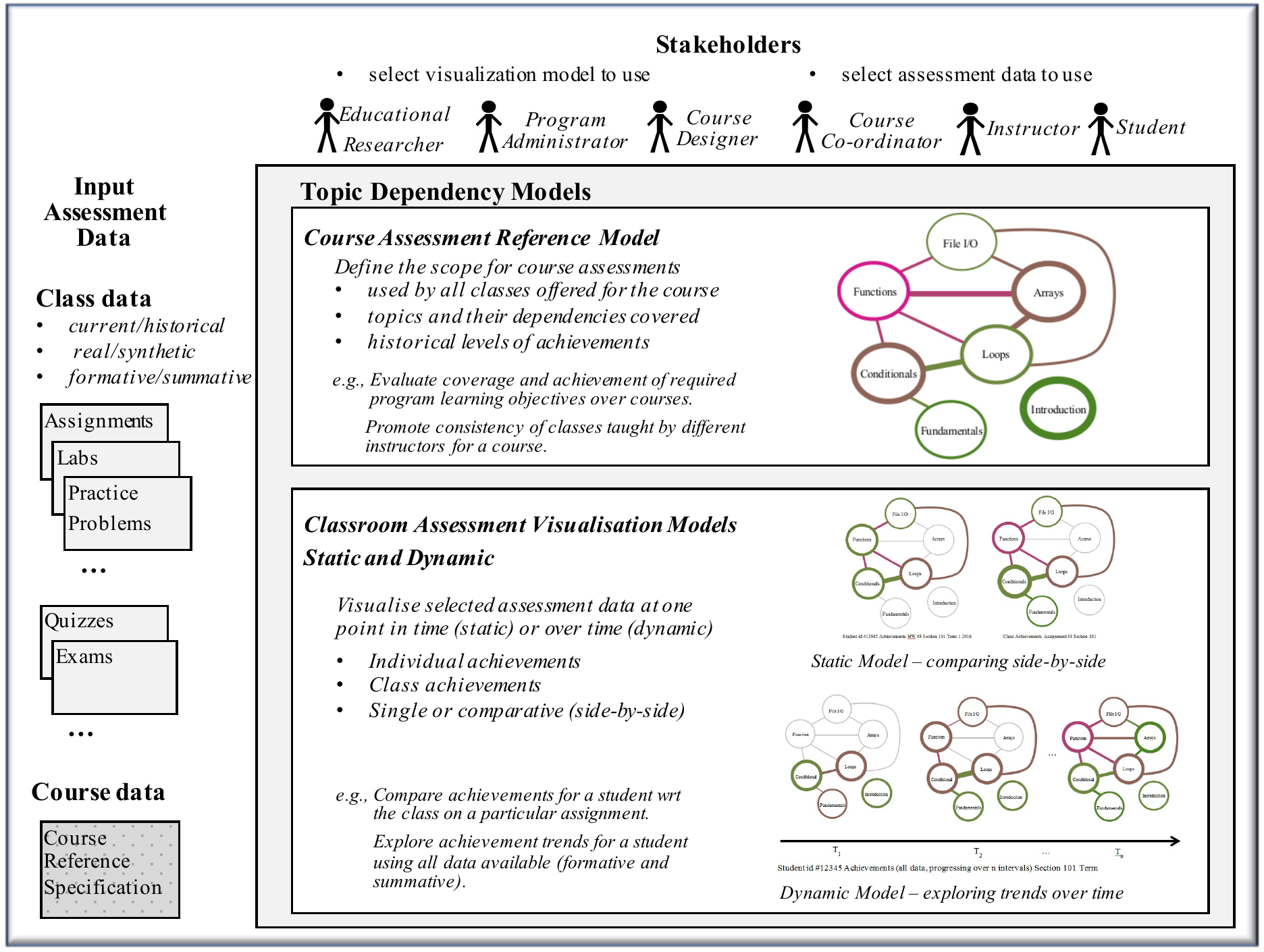}
        \caption{Original TDM model collection: two-weighted undirected graph foundation \cite{Khosravi2018}.}
        \label{fig:original-overview}
\end{figure*}

In this work, an alternative graph foundation is explored for the TDM collection, a two weighted hypergraph; the current results focus on the static classroom models. This graph can explicitly visualise the n-ary topic coverage and achievement inherent in questions (i.e., a question can assess 1:N topics). However, visualisations of their more general n-ary hyperedge relationships may be more difficult to understand in the broader community. Here, a method to generate a two-weighted hypergraph model using commonly available assessment data and a flexible, approach to address the complexity of the graphs are presented. The visualisation approach is a multilevel filtering approach that considers the hypergraph as a collection of levels. The content of these levels can be customised and presented according to user preferences. For example, instead of being presented with the complete hypergraph, the users can select topics, achievement or number of responses (e.g., maximum, minimum, range of values) to view part of it. In addition, the visualisations of the levels can be presented in either a cummulative or accumulative mode. The new models are presented using an illustrative example. A case study that illustrates how the model can be applied to provide insight for instructors in presented. This study uses historical data from a first-year undergraduate level offering of a course on programming and engineering design at The University of British Columbia. 


\section{Background: The Original TDM Collection}
A collection of TDMs has been presented in the author's previous works \cite{Khosravi2018, Cooper2018} to address some of the challenges stakeholders face that relate to the design, delivery and analysis of assessments. An overview of the collection is presented in Figure \ref{fig:original-overview}. 
\par{\bf{Stakeholders and Scenarios of Use.}} The wide variety of stakeholders interacting with these models (e.g., students, instructors, and so on) are shown at the top of the figure; they have roles inside and outside of the classroom.
The earlier results present a preliminary scenario analysis that identifies 40 questions for the stakeholders. For example, instructors interact with the models to explore questions such as: 
\begin{itemize}[itemsep=1mm, parsep=0pt]
    \item {What topics and their relationships do I need to assess?}
    \item{What topics are covered each of the assessments?}
    \item{How well are the students performing on the topics?}
    \item{How well are the students performing compared to other (current or previous) classes?}
    \item{Who in the class may be at risk of failing the course?}
    \item{How have all the topics and their relationships been assessed?}
    \item{How much has the class improved over time on the topics? }
\end{itemize}

\par{\bf{Input Data.}} The input assessment data, shown on the left of Figure \ref{fig:original-overview}, includes results from current or previous classes. The data can be from a variety of formative (e.g., assignments) or summative (e.g., quizzes) assessments; the course data includes a specification of the course topics that need to be assessed.

\par{\bf{Topic Dependency Models.}} Based on the requests from stakeholders, the input data are selected and transformed into visual models. As illustrated in main block of Figure \ref{fig:original-overview}, the TDM collection consists of two kinds of models: Course Assessment Reference and Classroom Assessment Visualisation (static and dynamic). The Reference model establishes and communicates the required topics and their relationships for a course (e.g., a CS1 course is an introductory programming course covering variables, branching, loops, and so on). It  provides a common foundation for all sections of a course offered over time (e.g., CS1 Section 001 Term 1 2019 Instructor  A. Smith). The reference model helps to ensure the consistent coverage of topics by different instructors, clearly communicate the scope of the topics to students, and support administrative activities related to monitoring learning objective outcomes.

The Classroom Assessment Visualisation models present assessment data within a class and support comparisons of assessment data between classes at one point in time (static) and over time (dynamic). The assessment data are selectively visualised, e.g., for the whole class, individual students, specific topics, and so on. For example, the user can choose to visualise static models, in a side-by-side comparison, for a specific assessment (e.g., Assignment 3) for all of the students in two classes. Alternatively, the user can choose to visualise dynamic models to explore the progression of a class over time. 

\par{\bf{Graph Foundation of the TDMs}} The Reference and Classroom Assessment Visualisation models share a common graph foundation: a two-weighted undirected graph. The vertices represent topics in a course; the edges represent assessment material (e.g., questions) that address the topics the edges are related to. The weights are reflected in the viusalisation using a colour palette (achievement) and width (coverage). As the graphs can only represent edges with up to two vertices, questions involving three or more topics must be redistributed in the models as combinations, in order to visualise them. 


\section {Related Work}
\label{related-work}
The learning analytics community continues to actively investigate approaches that support the exploration of learning activities by different stakeholders. With the increase in the use of educational technologies and the advancements in the areas of learning analytics and educational data mining, a new field, commonly known as ``Learning Dashboards" has emerged to help make sense of data sets in learning and education \cite{schwendimann2016perceiving,Kirschner2012,Park2015}. A variety of visualisations such as bar charts \cite{Dawson2012}, pie charts \cite{Tervakari2014}, histograms \cite{Leony2012}, box plots \cite{Albert2010}, radar graphs \cite{May2010}, and skill meters \cite{Bull2016} have been adopted to show the achievements of students for independent (stand-alone) topics.
\cite{schwendimann2016perceiving} presents a systematic literature review on the use of learning dashboards. Based on the findings of this literature review, the use of graph-based visualisations in learning dashboards has not received much attention. In addition, a number of studies provide strong evidence that opening the model to learners, leading to the notion of Open Learner Models (OLMs) \cite{bull2010open}, can be effective in helping students learn \cite{bodily2018open}. The OLMs commonly use a set of individual topics as their underlying structure for modeling learners' knowledge state, which ignores relationships among topics. An emerging new field applied methods from process mining and sequential data mining to educational data \cite{bogarin2018survey} to facilitate better understanding of the educational process. Educational process mining methods predominantly use graphs, but their main focus in on using activity logs to visualise students' learning process in terms of the time, place, path, pace of learning activities.

Hypergraphs have been adopted as a foundation for data analytics and visualisations in a wide variety of domains including data warehousing, communication network analysis, geospatial metadata, and cellular biology networks. To support interactive queries in data warehousing systems, a framework for developing dashboardscalled Dashboard-by-example has been proposed \cite{Hoang2012}. This framework adopts hypergraph-based techniques to transform dissimilar, heterogeneous data into a homogeneous knowledge space of clusters and partitions. The framework is flexible, as the hypergraph-guided data linkages support the exploration and aggregation of data from multiple perspectives. For the analysis of network traffic traces, available in massive communication logs, a hypergraph based visualisation is proposed in \cite{Glatz2014}. The network traffic visualization approach utilises a frequent item set mining method to identify interesting traffic patterns in the large amount of data. The patterns are visualised as hypergraphs with explicit, multi-attribute relationships. A framework to comprehensively address all of the available geospatial metadata standards (i.e., documents) is proposed in \cite{Rajaram2018}. Geospatial metadata describe geographic digital data resources such as earth imagery, geospatial databases and catalogues, and Geographic Information System files. The framework integrates hypergraphs and topic maps, representing the elements and their dependency relationships. The potential for adopting hypergraphs in the domain of cellular biology is introduced in \cite{Klamt2009}; recently, for example, \cite{Shen2018} proposes a framework that adopts hypergraphs and associated hypergraphs to describe, analyse, and identify metabolic network alignments at the full genome level. These alignments are used to discover important similarities and differences between distinct molecular networks: they reveal mappings between components (topological, biological functional) across different networks. 

To the best of our knowledge, a modeling and visualisation approach based on hypergraphs for educational assessment is not available at this time. The $TDM_{MH}$ helps to address this gap in the literature.   

\Section{Methodology}
This section define the graph foundation (hypergraph, level), scenarios of use, and an illustrative example. The graph generation and visualisation algorithm are presented in Section \ref{sec:generating-TDM}, using the example. A case study demonstrating $TDM_{MH}$s based on data collected from the final exam of a large introductory course on programming is presented in Section \ref{sec:case-study}.

\subsection{Hypergraph Foundation of the $TDM_{MH}$}
\label{sec:hypergraph}
In this section, the formal definition for a two-weighted hypergraph and the concept of a level are presented.

\begin{Definition} \label{def:TDP}
A two-weighted, undirected hypergraph $G = (V, H)$, where $V$ is the set of vertices representing the topics, and $H$ is the set of hyperedges. A hyperedge $h \in H$ is represented as $h= (C,c_1, c_2)$, where $C$ is the subset of the vertices being connected, and $c_1$ and $c_2$ represent the two weighted values for an edge. An edge involving only one node represents a self-loop.
\end{Definition}

\begin{figure}
		\centering
        \includegraphics[width=0.48\textwidth]{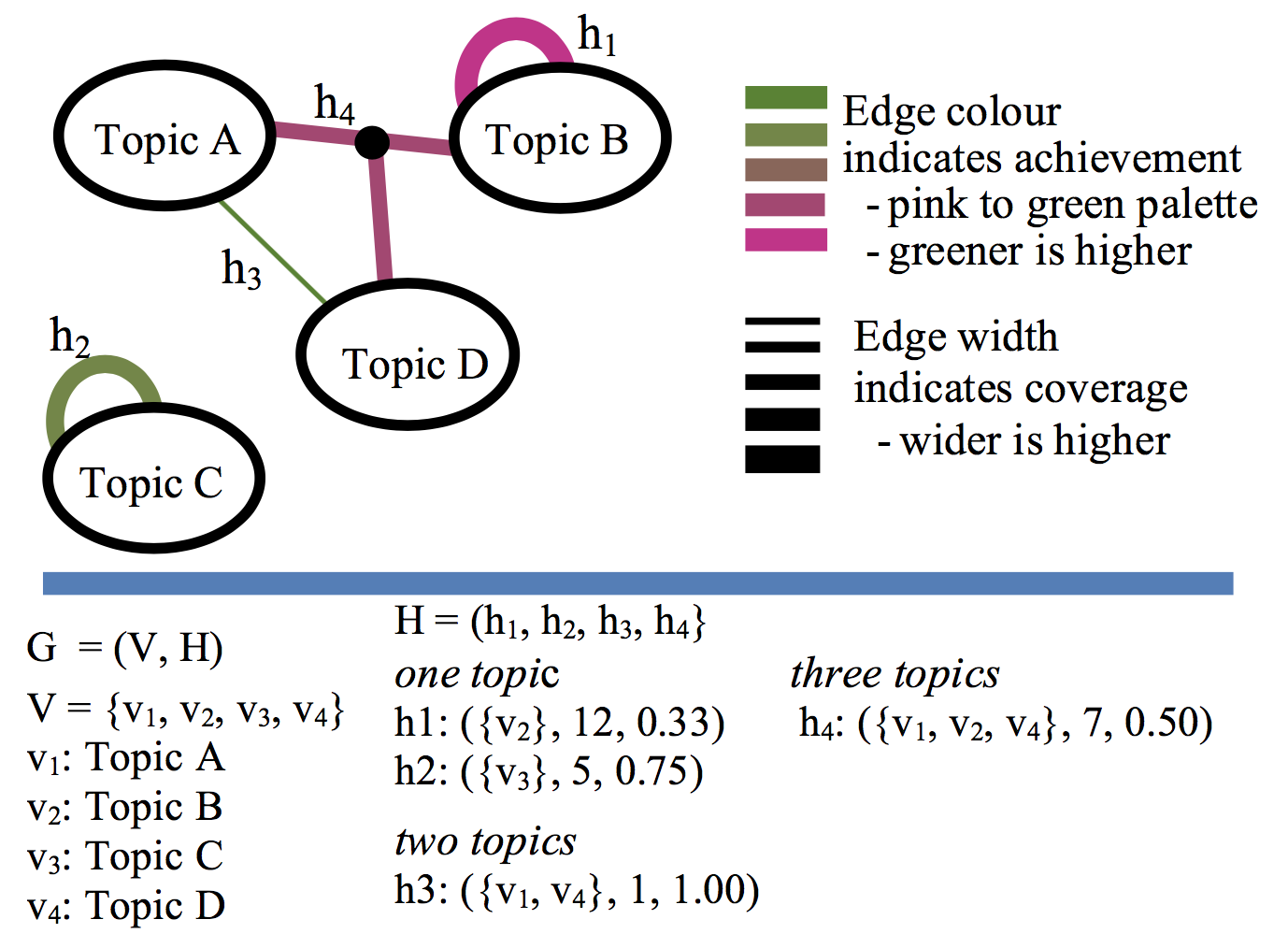}
        \caption{Example of a two-weighted hypergraph: definition and visualisation}
        \label{fig:hypergraph-overview}
\end{figure}

In this work, the nodes are used represent topics; the hyperedges represent the assessment material that covers the topics. The weight $c_1$  represents the number of learning objects that are tagged with the topics for the hyperedge; and $c_2$ represents the represents the performance (e.g., achievement) on these learning objects. A simple example of a $TDM_{MH}$ visualisation is illustrated in  Figure~\ref{fig:hypergraph-overview}. There are four topics in this example (A, B, C, and D); each topic is represented by a node. Topic B and Topic C both have a 1-ary hyperedge (i.e., a self loop). For Topic B, the edge has a dark pink colour and a wide width, reflecting the poor average performance (33\%) on the questions answered and the large number of responses. The hyperedge for Topic C is a medium green colour and has a medium width; this represents the good average performance (75\%) and a moderate number of responses to questions.  A 2-ary hyperedge exists between Topics A and D. This edge is dark green and thin, in representing the excellent performance (100\%) on the related questions and a small number of responses. A 3-ary hyperedge exists between Topics A, B, and D. This edge is a brownish pink colour and has a moderate weight, which represents the moderate average performance (50\%) and a moderate number of responses to questions on these three topics. 

The concept of a level is used in this work to support the visualisation. For example, in the simple example above, there are three levels. $Level_1$ is a subset of the hyperedges with one vertex (i.e., $h_1$, $h_2$), $Level_2$ is a subset of the hyperedges with two vertices (i.e., $h_3$), and $Level_3$ is a subset of the hyperedges with three vertices (i.e., $h_4$). More generally, $Level_i$ is the subset of hyperedges involving i vertices.

\subsection{Scenarios of Use}
As with the orignal TDM models, the users can select the assessment data of interest. For example, they may need to explore data for formative or summative assessments (e.g., one or more assignments or examinations); the data can be from one or more classes. Here, scenarios of use include filtering by topic(s), or constraints involving achievement or coverage. Additional filters can be added in a straightforward way. Currently, the scenarios focus on using the static models; the reference and dynamic models for the $TDM_{MH}$ are planned in future work. Two examples, derived from the scenario analysis in \cite{Khosravi2018}, are described below.  

\textbf{Topics}. How are the students doing on a specific topic? One or more topics of interest can be selected in order to identify topics that may need additional attention. For example, an instructor can select one of many topics covered in a course; the levels are filtered to emphasize the topic(s) of interest.  

\textbf{Achievements}. What topics have poor achievement? A constraint for visualising a particular achievement level can be selected. For example, an instructor can select to view the minimum achievement scores over all topics in order to identify those that need additional attention; the levels are filtered to emphasize the achievement scores of interest.


The users can choose to combine filters. In other words, they can filter the visualisations based on one or more constraints such as selecting both a topic and an achievement constraint, e.g. highest performance involving topic $T_1$. The results are presented on a level-by-level basis (a level is described in Section \ref{sec:hypergraph}.

When visualising the results, the user can also choose to view the results in either the cummulative or accumulative mode.  In the accumulative mode, only the results for one level are presented. In the cummulative mode, the results up to and including the current level of interest are presented.

\subsection{Illustrative Example}
An example based on six students, three formative assignments (five questions per assignment), and six topics has been defined by the authors. The following constraints are considered for creating the illustrated example:
\begin{itemize}[itemsep=1mm, parsep=0pt]
    \item The questions must span a range of one to four topics.
    \item There must be a range in the individual student achievements in the responses (low to high scores).
    \item There must be a range in the number of responses to the questions. 
    \item There must be a range in the average scores for the questions (0\%-100\%).
    \item Assessments on groups of topics demonstrate a range of achievements (very poor to very strong); one or more questions address the groups of topics.
    \item The multilevel visualisation algorithm with a list of filtering options (e.g., selected topic(s), achievement, cumulative/accumulative mode) can be clearly illustrated.
\end{itemize}
 
 Table~\ref{tab:student-response-data} presents a sample data set meeting these constraints. As input to the $TDM_{MH}$ generation algorithm, the data are formatted into two CSV files: (1) A student-question-answer (SQA) file that contains the student identifier, question identifier, and score (correct/incorrect), and (2) a question-topic (QT) file that contains the question identifier and topics (i.e., tags) it addresses. 

\begin{table}[t]
\centering
\footnotesize{
\begin{tabular}{|l||l||llllll|}
\hline
Q & Topic List & \multicolumn{6}{|l|}{Student ID }\\

 &  & $S_1$ & $S_2$ &$S_3$ &$S_4$ &$S_5$ &$S_6$ \\
\hline \hline
\multicolumn{8}{|l |}{\textit{Assignment 1}} \\
\hline
$Q_1$ &  $T_1$              & - & 1 & - & - & 0 & 1  \\
$Q_2$ & $T_3$              & 1 & 1 & 1 & 1 & 1 & 1     \\
$Q_3$ &  $T_4$              & 0 & 1 & 0 & 1 & 1 & 1     \\
$Q_4$ &  $T_1$, $T_2$          & 1 & 1 & 0 & 1 & 1 & 0     \\
$Q_5$ &  $T_1$, $T_4$          & 1 & - & - & 0 & 1 & 0   \\
\hline
\multicolumn{8}{|l |}{\textit{Assignment 2}} \\
\hline
$Q_6$ &  $T_4$             & 0 & 0 & 1 & 0 & 0 & 1      \\
$Q_7$ &  $T_1$, $T_4$          & - & 0 & 1 & 1& 0 & 0      \\
$Q_8$ &  $T_4$, $T_5$          & 0 & 1 & 0 & 0 & 0 & 1      \\
$Q_9$ &  $T_1$, $T_4$, $T_5$       & 0 & 0 & 0 & 0 & - & 1     \\
$Q_{10}$ &  $T_1$, $T_2$ ,$T_4$, $T_5$   & 0 & 0 & 0 & 1 & 0 & 1      \\
\hline
\multicolumn{8}{|l |}{\textit{Assignment 3}} \\
\hline
$Q_{11}$ &  $T_1$, $T_4$          & 1 & - & 0 & 1 & - & 1  \\
$Q_{12}$ &  $T_2$, $T_6$          & 0 & 0 & 0 & 0 & 0 & 0     \\
$Q_{13}$ &  $T_1$, $T_2$, $T_6$     & - & - & 1 & 1 & - & 1 \\
$Q_{14}$ &  $T_1$, $T_2$, $T_4$, $T_5$    & - & 0 & 1 & 0 & 0 & 1   \\
$Q_{15}$ &  $T_2$, $T_4$, $T_5$, $T_6$   & - & 0 & 0 & 1 & 1 & 1   \\
\hline
\hline
\end{tabular}
}
\caption{Illustrative example: questions, topic lists, and student responses. The value 1 indicates the question is answered correctly; 0 indicates it is answered incorrectly; and - indicates it is not attempted.}
\label{tab:student-response-data}
\end{table}

\section{Generating and Visualising $TDM_{MH}$: \\Classroom, Static Model} 
\label{sec:generating-TDM}

In this section, the approach to generating and filtering a $TDM_MH$ is presented. Section~\ref{sec:generate-TDMs} presents methods for generating the graph and Section~\ref{sec:multilevel-visualisations} explores filtering mechanisms to address the complexity of hypergraph visualisations. 


\subsection{Generating $TDM_{MH}$}
\label{sec:generate-TDMs}
This section demonstrates how commonly available input data (student achievements/grades for specific questions) and the mapping from the questions to the course topics are transformed into a $TDM_{MH}$.  High-level code and notation are presented in Algorithm~\ref{alg:TDM-Static}.
The algorithm consists of three high level steps: create working dictionaries and matrices; define the $TDM_{MH}$ graph elements (vertices and edges); and visualise (i.e., plot) the $TDM_{MH}$ graph. 

\begin{algorithm}
  \caption{Generating a $TDM_{MH}$}
  \label{alg:TDM-Static}
  \begin{algorithmic}[1]
  			\Require $SQA.CSV$, $QT.CSV$, $filters$
  			\Statex{\textbf{Create dictionaries and matrices for efficient indexing}}
  			\State $QDict \gets CreateQDict (SQA.CSV)$ 
  			\State $SDict \gets CreateSDict (SQA.CSV)$
  			\State $TDict \gets CreateTDict (QT.CSV)$
  			\State $T \gets CreateT (QT.CSV, QDict, TDict)$ 
  			\State $A \gets CreateA (SQA.CSV, SDict, QDict)$
			\State $D \gets CreateD (SQA.CSV, SDict, QDict)$
			\Statex{}
  			\Statex{\textbf{Compute the Graph Elements: Vertices and Edges}}
  			\State $VList \gets ComputeV (TDict)$   
  			\State $HList \gets ComputeH (T, A, D, TDict)$ 

  			\Statex{}
  			\Statex{\textbf{Create and visualise the Graph}}
  			\State $TDMStatic \gets CreateTDM(VList, HList)$
  			\State $Visualise(TDMStatic, filters)$
  \end{algorithmic}
\end{algorithm}

The first six steps of Algorithm \ref{alg:TDM-Static} transform the data in the SQA.csv and QT.csv files into working dictionaries and matrices. In Steps 1, 2, and 3 of the algorithm, three dictionaries are created, $QDict$, $SDict$,  $TDict$, to map array and matrix indices to question, student, and topic identifiers. In Steps 4, 5, and 6, three working matrices are created: T, A, and R. The information on topics assigned to each question is represented in matrix $T$, in which $t_{ij}=0$ indicates that question $i$ is not tagged with topic $j$ and $t_{ij}=1$ indicates that question $i$ is tagged with topic j. 
The correctness of the answers provided by the users are represented in matrix $A$, where $a_{ij} = 0$ indicates that user $i$ has answered question $i$ incorrectly,  $a_{ij} = 1$ indicates that user $u$ has answered question $i$ correctly, and  $a_{ij} = n$ indicates that user $u$ has not attempted question $i$. Matrix $R$ is used to keep track of attempted questions, where $r_{ij}= 0$ if user $i$ has not attempted question $i$ and one otherwise. 

Steps 7 and 8 in Algorithm \ref{alg:TDM-Static} establish the graph vertices and hyperedges. $VList$ stores the list of the vertices of the $TDM_{MH}$ graph and $HList$ stores the list of its hyperedges. The coverage and competency associated with an edge are both computed using $T$, $A$ and $R$ within the $ComputeH$ function. The coverage ($Cov$) associated with a hyperedge among a set of vertices $V={v_j, ..., v_k}$ is computed using the following formula:

\begin{equation}\label{formula:covereage}
Cov(V) = \sum_{i \in QDict} t_{ij}\times ... \times t_{ik} \sum_{u \in SDict} d_{ui}
\end{equation}

The outer summation loops through all of the questions. A question $i$ contributes to $cov(V)$ only if it is tagged with exactly the topics that are included in $V$ (i.e., $t_{ij} \times... \times t_{ik} = 1$). For example, a question tagged with topics ({$T_1$, $T_2$, $T_3$}) attempted by nine learners  contributes $\frac{9}{1+0} = 9$ to the coverage of $Cov(\{T_1, T_2, T_3\})$. 

The achievement ($Achv$) associated with a hyperedge between a set of vertices  $V={v_j, ..., v_k}$ is computed using the following formula:

\begin{equation}\label{formula:competency}
\centering
Achv({V}) = \frac{\sum_{i \in QDict} t_{ij}\times ... \times t_{ik} \sum_{u \in SDict}   a_{ui}}{Cov({V})}
\end{equation}

A question $i$ contributes to $Achv({V})$ only if it is tagged with exactly the topics that are included in $V$(i.e., $t_{ij} \times ... \times t_{ik} = 1$). The extent of the contribution is determined by the the number of learners who have correctly answered the question, which is computed by $\sum_{u \in SDict} a_{ui}$. The numerator of this formula computes the number of times questions tagged with exactly topics that are included in $V$. Dividing this number by $Cov(V)$ produces the probability of correctly answering questions on topics in $V$. Note that $Cov$ and $Achv$ of self-loops can also be computed via the same two formulas by using $V = {v_j}$.

\begin{figure}[!htb]
		\centering
        \includegraphics[width=0.47\textwidth]{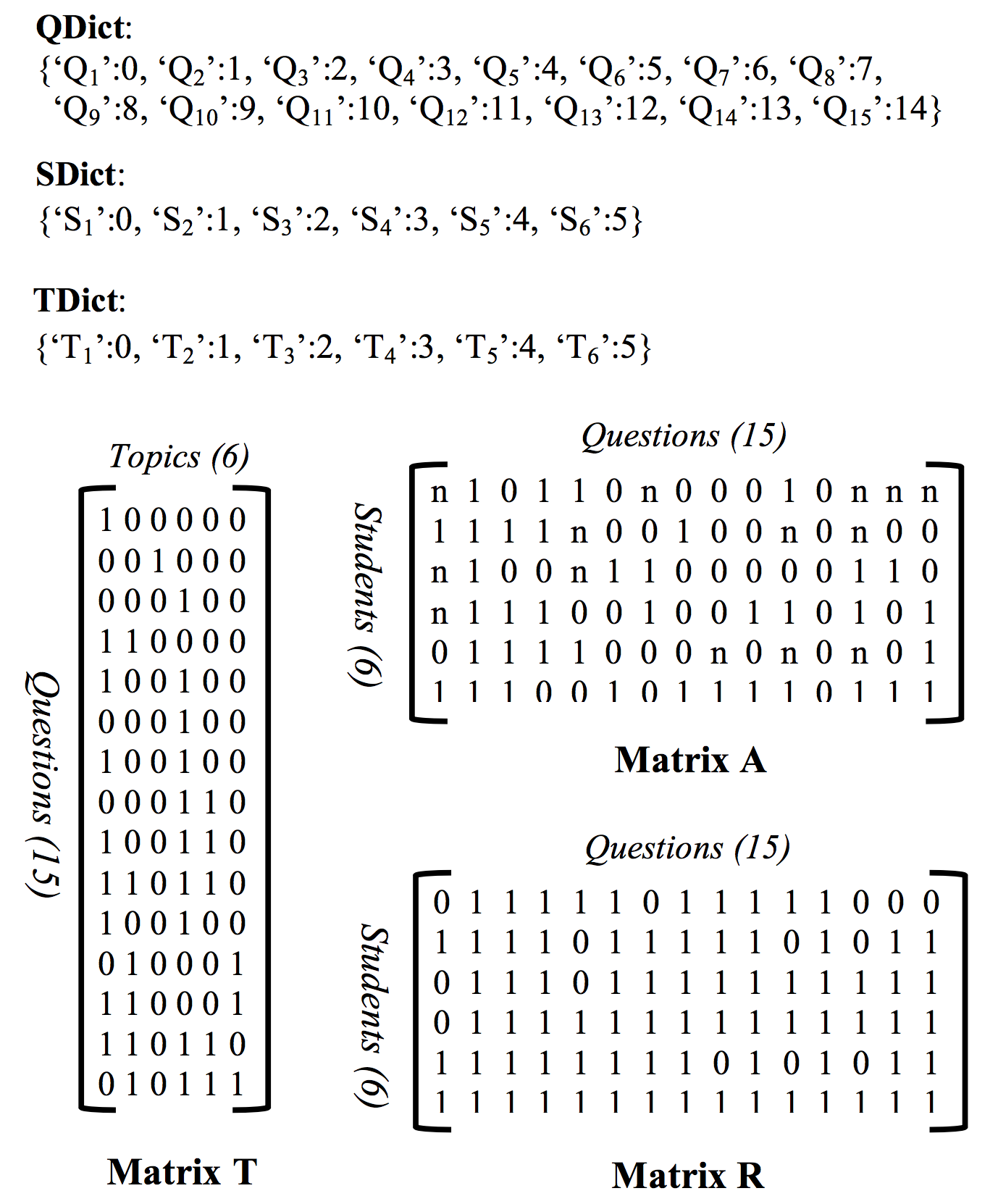}
        \caption{The data represented in Table~\ref{tab:student-response-data} are organized into two input files: SQA.csv and QT.csv. The data are loaded, transformed into three dictionaries: $QDict$, $SDict$ and $TDict$ (Steps 1, 2 and 3 in Algorithm~\ref{alg:TDM-Static}). The dictionaries and the input files are used to create three working matrices: T, A, and R (Steps 4, 5 and 6 in Algorithm~\ref{alg:TDM-Static}).
        These matrices are used to create the $TDM_{MH}$ model.}
        \label{fig:example-input}
\end{figure}

Figure~\ref{fig:example-input} represents the data which are stored in dictionaries ($QDict$, $SDict$, $TDict$) and matrices ($T$, $A$, $R$). 

\begin{figure}[!htb]
		\centering
        \includegraphics[width=0.5\textwidth]{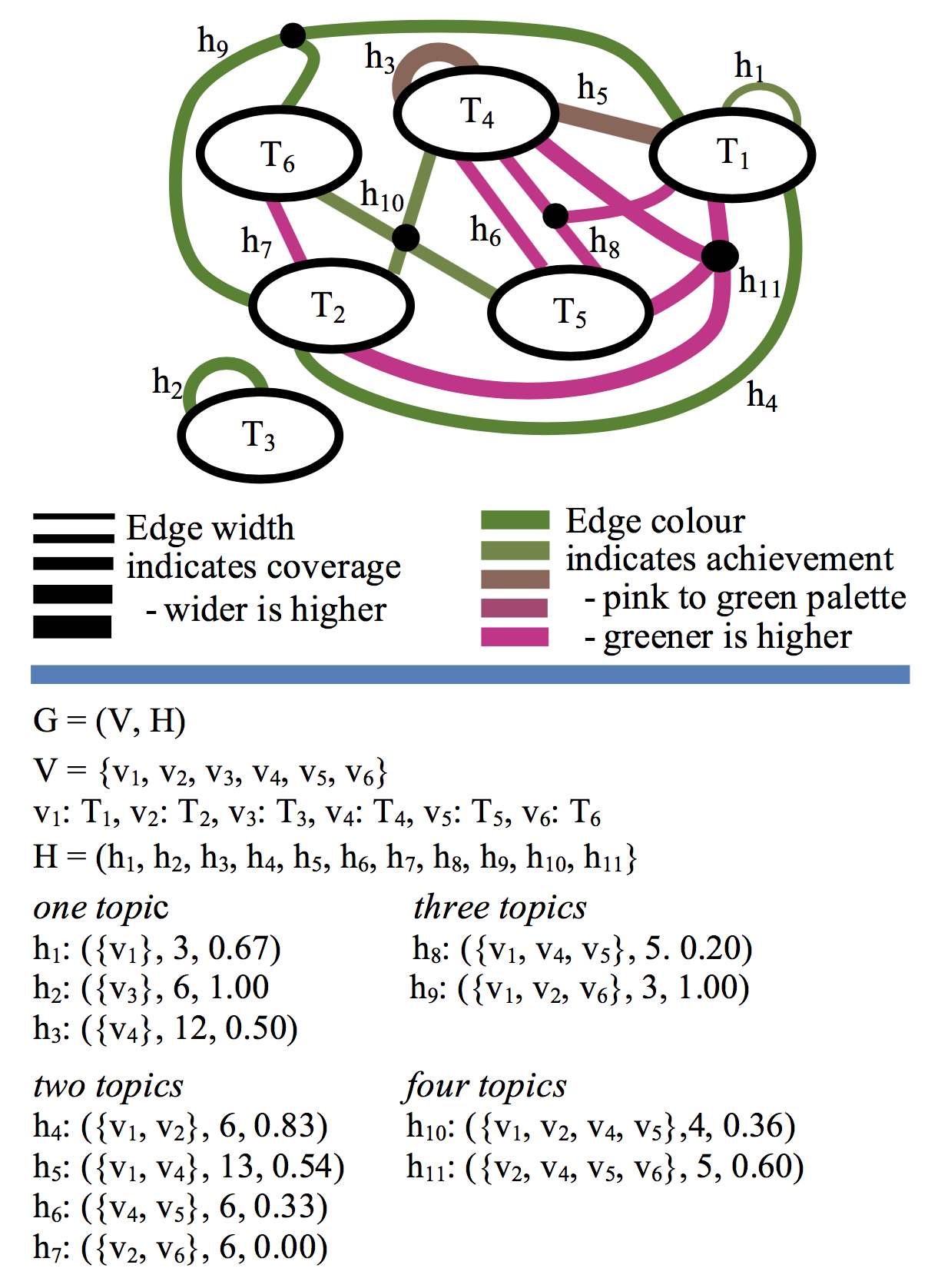}
        \caption{Complete $TDM_{MH}$ for the data set provided in Table~\ref{tab:student-response-data} using Algorithm~\ref{alg:TDM-Static} (no filtering). The example does not have questions involving five or six topics.
        \label{fig:example-Static}}
\end{figure}

Steps 9 and 10 of Algorithm \ref{alg:TDM-Static} create and visualise the graph. Figure~\ref{fig:example-Static} illustrates the resulting model, without any filtering. It has six vertices as the questions in this data set are tagged with six topics: $T_1$, $T_2$, $T_3$, $T_4$, $T_5$, and $T_6$. The coverage of the hyperedges are computed using Formula~\ref{formula:covereage}. For example, the hyperedge with id $h_5$ has coverage of 13, which consists of contributions of 4 from question 5, 5 from question 7, and 4 from question 11. The achievement of the hyperedges are computed using Formula~\ref{formula:competency}. For example the achievement of the hyperedge with id $h_5$
is computed as $\frac{7}{13}=0.54$.

\subsection {Visualising $TDM_{MH}$}
\label{sec:multilevel-visualisations}
\begin{figure*}[!htb]
		\centering
        \includegraphics[width=1.0\textwidth,height=5cm]{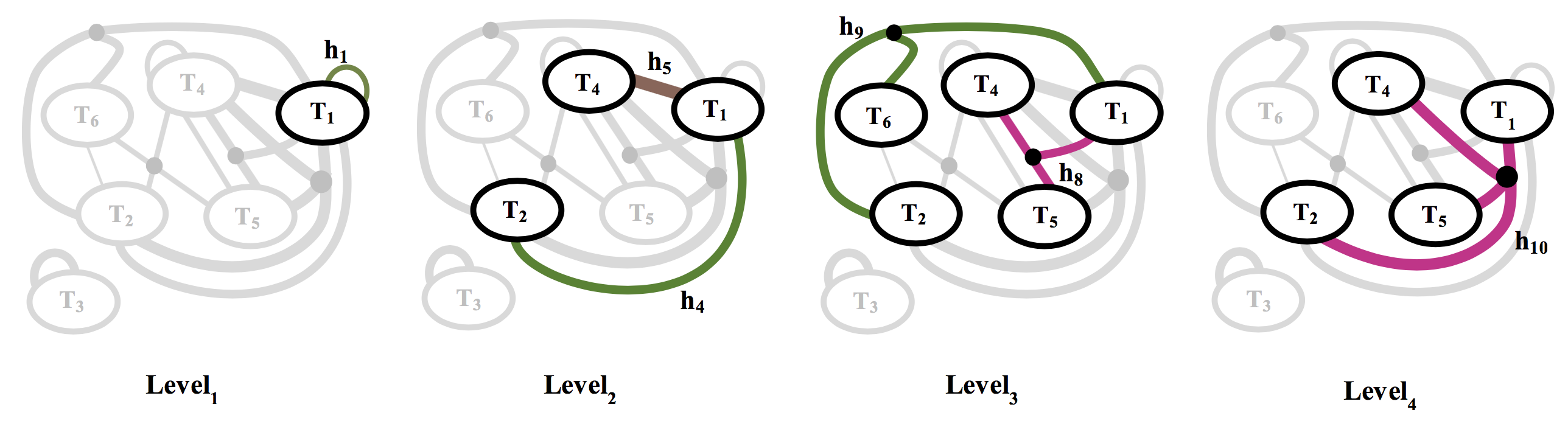}
        \caption{Multilevel visualisation: one topic is selected ($T_1$) as a filter (accumulative mode). }
        \label{fig:multilevel-case1}
\end{figure*}


In Step 10 of Algorithm \ref{alg:TDM-Static}, the $TDM_{MH}$ can be presented in either whole or part according to the filters the user selects. To address the complexity of the model, the topology of the graph is used; the illustrative example (with six nodes) provides six levels. Table \ref{tab:filtering-example} illustrates the levels before any filtering is done. For example, $Level_1$ is the subset of hyperedges \{$h_1, h_2, h_3$\}.

If the user does not select any filters, then all levels are displayed by default, including all topics, achievements, and coverage of the learning objects as illustrated in Figure~\ref{fig:example-Static}. Alternatively, the selected filters are applied to each level. The selected elements are presented using the original TDM colour palette; other items are greyed out. $Level_i' \gets filter(Level_i, t, a, n)$, where the parameters t, a, and n represent optional constraints for the topics, achievements, and number of learning objects. For example, if a user requests to filter with respect to topic $T_1$, the resulting filtered levels are illustrated in Table \ref{tab:filtering-example}. In this case, for example, the filtered $Level_i'$ is the subset of hyperedges \{$h_1$\}, as only $h_1$ contains $T_i$.

After the levels are filtered, they are ready to be visualised. The user can choose to present them in either an accumulative or cumulative mode. In the accumulative mode, only the hyperedges for one level are presented at a time. The first four levels are illustrated in \ref{fig:multilevel-case1} in the accumulative mode. In the cumulative mode, the presentation for a level includes the hyperedges for the current levels and the previous levels. A preliminary layout algorithm is used in the prototype tool at this time; however, adopting a more effective proposal from the literature is planned in the next step. 

This straightforward visualisation example based on filtering the levels for one topic can be extended to combinations of one or more topics, achievements, and coverage. The options to highlight part of the graph, in either cumulative or accumulative mode, provide a flexible approach for users. However, to easily access more detailed information (e.g., exploring specific achievement values for hyperedges), additional research is needed to extend the $TDM_{MH}$ visualisations. 

\begin{table}[t]
\centering
\footnotesize{
\begin{tabular}{|p{1.1cm}|p{2.2cm}|| p{2.3cm}|}
\hline
Level    & Level & Level' \\
ID       & Membership & Membership ($T_1$)
\\
\hline
\hline
$Level _1$ & \{$h_1, h_2, h_3$\} & \{$h_1$\}
\\
\hline
$Level _2$ & \{$h_4, h_5, h_6, h_7$\} & \{$h_4, h_5$\}
\\
\hline
$Level _3$ & \{$h_8, h_9$\} & \{$h_8, h_9$\}
\\
\hline
$Level _4$ & \{$h_{10}, h_{11}$\} & \{$h_{10}$\}
\\
\hline
$Level _5$ & $\emptyset$ & $\emptyset$
\\
\hline
$Level _6$ & $\emptyset$ & $\emptyset$
\\
\hline
\hline
\end{tabular}
}
\caption{Levels in the illustrative example graph: before and after filtering (e.g., topic $T_1$).}
\label{tab:filtering-example}
\end{table}





\section{Case Study}
\label{sec:case-study}

\begin{figure*}[!htb]
		\centering
		\includegraphics[width=14.5cm]{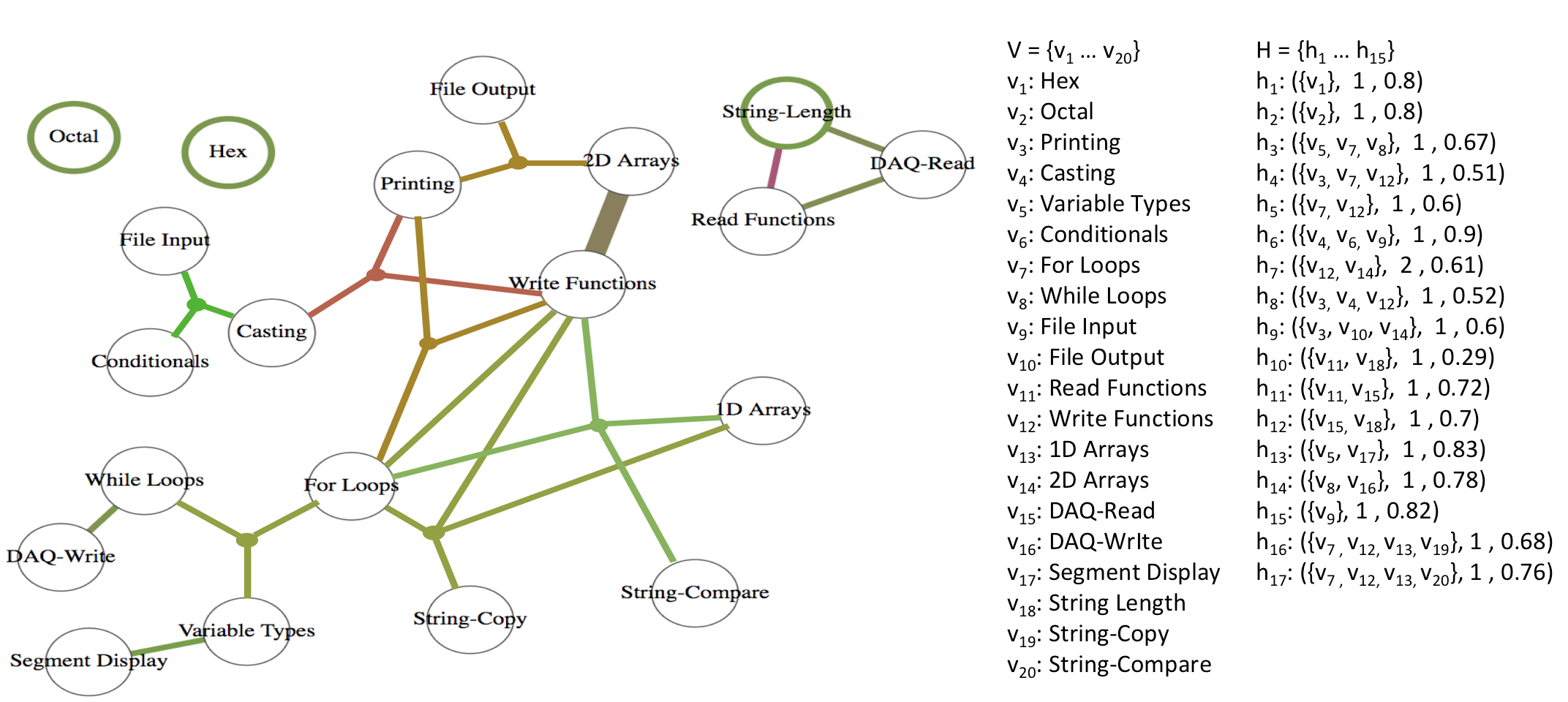}
        \caption{$TDM_{MH}$ for the final exam of an offering of a first-year course on programming and engineering design}
        \label{fig:APSC160-exam-module}
\end{figure*}
This section presents a case study that demonstrates the applications of the $TDM_{MH}$. This case study is based on data collected from a first-year undergraduate level offering of a course on programming and engineering design at The University of British Columbia. This offering of the course had 377 students and was held during the Fall of 2016. The course covers many topics that are generally included in an introductory course on programming and engineering design in eight modules: number conversions, programming fundamentals, conditionals, loops, file I/O, functions, arrays, strings, and DAQ systems. Functions, strings and DAQ systems received two weeks of lecture time; all of the other modules received roughly one week of lecture time. 

We use data collected from the final exam of the course, which is captured captured via the Gradescope platform \cite{singh2017gradescope}, a system for the on-line assessment of handwritten exams, are used. The final exam of this offering had 17 independent sub-questions that formed a total of eight main questions. The questions are tagged using 20 concept-level topics, defined by the instructors, based on the eight modules that are covered in this course. For example, the Strings Module is further decomposed into three topics: String-Length, String-Copy, and String-Compare, providing a finer level of granularity. 

Figure~\ref{fig:APSC160-exam-module} shows the $TDM_{MH}$ for this exam. It indicates that the exam adequately covers all of the modules of the course and that all modules except Conversions (including the Hex and Octal topics) are highly connected to one another. The exam includes questions that had only a single topic (e.g., $h_1$ on Hex), two topics (e.g., $h_{14}$ on While loops and DAQ-write), three topics (e.g., $h_9$ on Printing, File output and 2D arrays) and four topics (e.g., $h_{17}$ on For loops, 1D arrays, Write functions, and String-compare).

The further decomposition to concept-level tags provides insights which might have not been possible to gauge using module-level tags. For example it shows that students have done well on questions on String-Copy and String-Compare but not so well on questions on String-Length. As another example, further decomposition of items from the Arrays Module shows that students are able to do well in 1D-Arrays, but perform quite poorly on questions on 2D-Arrays. 

\begin{figure}[!htb]
		\centering
		\includegraphics[width=8.5 cm]{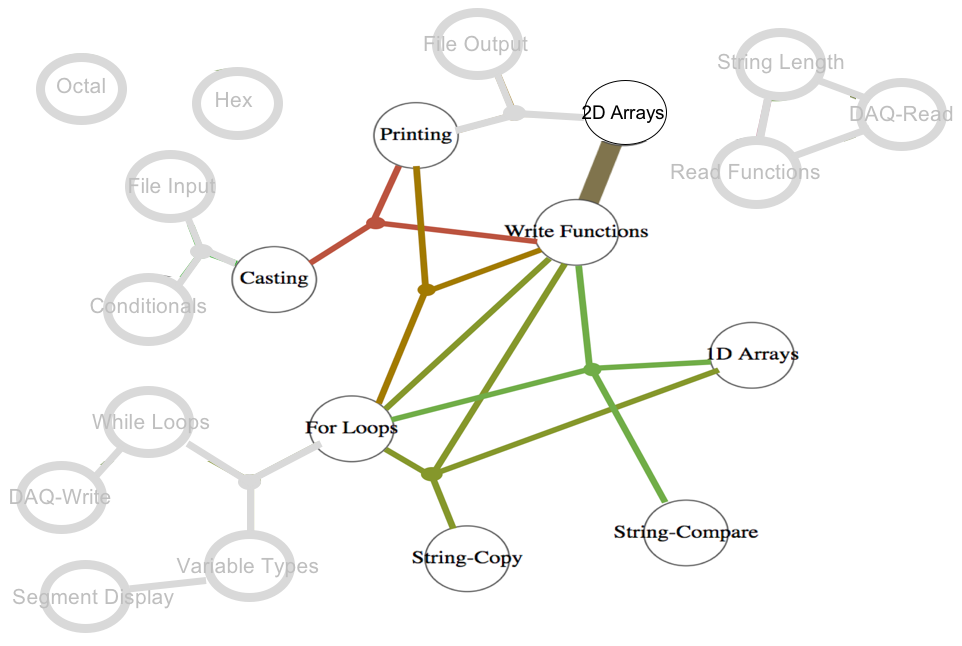}
        \caption{Cumulative topic-based $level_4$ filter on ``Write Functions" is selected.}
        \label{casestudy-filter1}
\end{figure}

Figure~\ref{casestudy-filter1} applies the cumulative topic-based $level_4$ filter on the ``Write Functions" concept. This filter will enable instructors to determine the coverage and performance of students on a particular concept. The $TDM_{MH}$ in the given example demonstrates that the write function has been extensively covered in the exam. The performance of students on questions that covered write functions in combination with 1D arrays, String-copy, and String-compare have been quite good; however,  their performance on questions that the cover the write functions in combination with 2D arrays, Casting, and Printing has not been as good. 

\begin{figure}[!htb]
		\centering
		\includegraphics[width=8.5 cm]{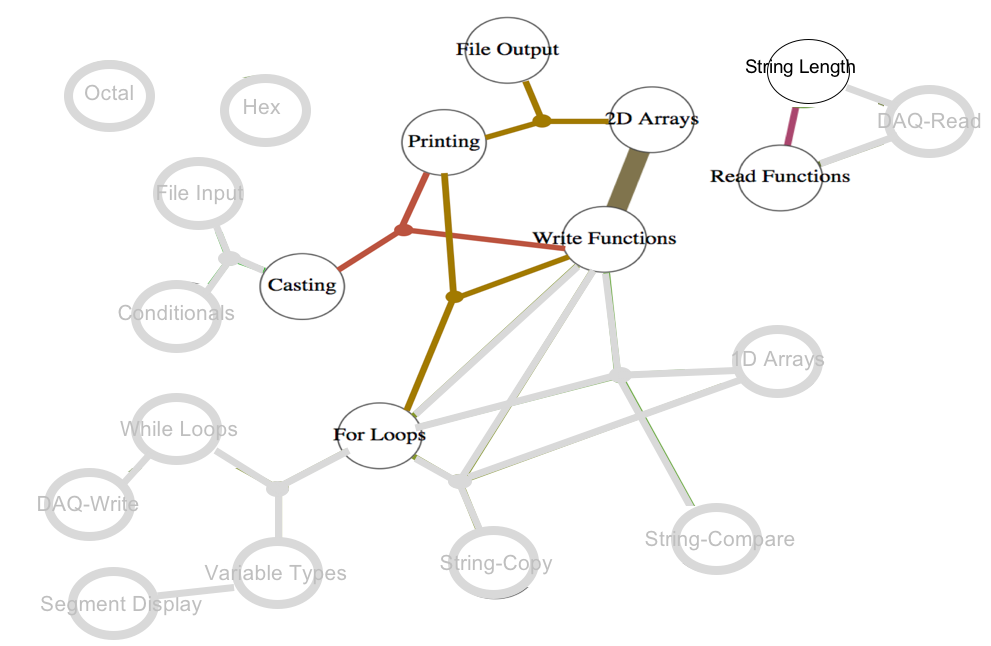}
        \caption{Cumulative achievement $level_3$ filter with ``$\leq 60$" is selected.}
        \label{casestudy-filter2}
\end{figure}

Figure~\ref{casestudy-filter2} applies the cumulative achievement $level_3$ filter with ``$\leq 60$". This filter enables instructors to determine gaps in students' knowledge. The $TDM_{MH}$ in the given example demonstrates questions concept combinations that the students have performed poorly on in this exam. Interestingly, all of the hyperedges that include Printing are selected in this filter. This may suggest that there is a general misunderstanding or misconception about how print statements are used. 

\section{Conclusions and Future Work}
The preliminary results of investigating an alternative graph foundation, a two-weighted hypergraph, for a collection of TDMs are reported in this work; the new model is called $TDM_{MH}$. The generation and visualisation of $TDM_{MH}$ are presented. The generation of the model utilises matrix computations, which makes it scalable and efficient. The visualisation helps to address the complexity of a hypergraph through a flexible, multilevel approach. A case study that illustrates how $TDM_{MH}$ can be applied to provide insight in the context of a large university course is presented.    

A number of directions are planned for the next steps in this research, including exploring the $TDM_{MH}$ dynamic model, refining the prototype tool, and additional validation studies, in particular a user study to compare the strengths and limitations of the original TDM and $TDM_{MH}$. In addition, highly interactive 3D visualisations of the $TDM_{MH}$ may also be of great value to the diverse stakeholders who are involved with the design and delivery of assessments. 
\small{
\bibliographystyle{elsarticle-num}
\bibliography{main}
}
\end{document}